\documentclass[12pt]{article}
\usepackage{amssymb}
\usepackage{graphicx}

\newcommand{\ee}{\end{equation}}
\newcommand{\be}{\begin{equation}}
\newcommand{\ba}{\begin{array}}
\newcommand{\ea}{\end{array}}
\newcommand{\bea}{\begin{eqnarray}}
\newcommand{\eea}{\end{eqnarray}}
\newcommand{\vev}[1]{\langle #1\rangle}

\newcommand{\preprintno}[1]
{\vspace{-2cm}{\normalsize\begin{flushright}#1\end{flushright}}\vspace{1cm}}

\title{\preprintno{\bf MCTP-01-47}
Breaking CP and supersymmetry with orbifold moduli dynamics}
\author{Thomas Dent \\{\em Michigan Center for Theoretical Physics, 
Randall Lab.,} \\
{\em University of Michigan, Ann Arbor, MI 48109-1120}}
\date{October 2001}

\begin{document}
\maketitle

\begin{abstract}
We consider the stabilization of string moduli and resulting soft super\-symmetry-breaking 
terms in heterotic string orbifolds. Among the results obtained are: formulae for the scalar 
interaction soft terms without integrating out the hidden sector gaugino condensate, which 
reduce to standard expressions in the usual ``truncated'' limit; an expression for the modular 
transformation of $A$-terms; a study of the minima of the scalar potential in the 
K{\" a}hler modulus direction; and a discussion of the implications for CP violation 
phenomenology.

Some closely related results have appeared in a recent paper of Khalil, Lebedev and Morris, 
namely, the exact modular invariance of $A$-terms up to unitary mixing, and the existence 
of certain complex minima for string moduli. 
\end{abstract}

\section{Introduction}
Many phenomenologically important quantities in the low-energy effective theory of heterotic 
string orbifolds, including gauge and Yukawa couplings, depend on the vacuum expectation 
values (v.e.v.'s) of {\em moduli}\/. These are scalar fields that parameterize the metric 
and background fields of the compact space: their potential vanishes in perturbation theory, 
so it is essential to have some mechanism for stabilizing their values. Similar remarks apply 
to Wilson lines that can take a continuous range of values. Once supersymmetry (SUSY) 
is broken, necessarily by nonperturbative effects, the moduli get, in general, 
a nonvanishing scalar potential, or equivalently the compactification parameters are 
dynamically determined \cite{Kikkawa:1984cp}. If supersymmetry is broken at an energy scale 
far below that of the string, one should be able to study modulus stabilization using an 
effective field theory (see {\em e.g.}\/\ \cite{FontILQ90,FerraraMTV}).

We will focus on the implications of the moduli dynamics for CP violation, and other issues
of phenomenology, in the resulting low-energy theory of softly-broken supersymmetry. 
The resulting phenomenology is in general described by complex Yukawa couplings 
$y^{u,d,e}_{ij}$ and soft terms \cite{DuganGH84}, 
namely the gaugino masses, bilinear Higgs couplings $B\mu H_U H_D$ and trilinear scalar 
couplings (``$A$-terms'') $A^u_{ij} y^u_{ij} \tilde{q}_i \tilde{u}^c_j H_U$ {\em etc.}.
The phases of these terms are tightly constrained by experimental limits on electric dipole 
moments (for flavour-diagonal interactions) 
\cite[and references therein]{Pokorski:2000hz,Falk:1999tm,Abel:2001vy}
while flavour-off-diagonal interactions in the quark mass basis are constrained by 
experimental results for $K$ and $B$ mesons \cite{Gabbiani:1996hi}.

Whilst studies have appeared parameterizing the contribution of the moduli and dilaton (a 
scalar field giving the value of the string coupling) to CP violation in soft terms 
\cite{Brhlik:2000pw} by various assumptions about the 
magnitudes and phases of $F$-terms, our work addresses the question of how these terms, and 
the sources of CP violation in Yukawa couplings, are to be generated from a consistent 
model of SUSY-breaking and modulus stabilization. Research along similar lines has been 
pursued by Bailin {\em et al.}\/\ 
\cite{Bailin:1997fh,Bailin:1998iz,Bailin:1998xx,Bailin:2000ra}.

The effective supergravity theory used here is a slight simplification relative to 
detailed derivations undertaken for specific orbifold models, for example we use the
overall modulus approximation and ignore other potential sources of CP violation in looking 
at one $T$ modulus only. Also, exact target-space 
modular invariance is an important part of our analysis, but in realistic models it might be 
modified or explicitly broken. It is relatively simple to go from the overall modulus
case to that of 3 ``diagonal'' moduli for the complex planes of an orbifold. The moduli may 
behave differently, but looking at them one by one the dynamics are similar to the overall
modulus case (although with different parameters): see \cite{Bailin:1998iz}, where in most 
of the explicit orbifold models it was found that one or two moduli (for which modular 
invariance was realized) could be identified as sources of CP violation and could be described 
essentially in the same way as an overall modulus. We believe that the analysis presented
here gives a reasonable sample of the behaviour of $T$-moduli to be expected in the many 
possible orbifold models.

There exist other potential sources of CP violation in heterotic string models,
for example twisted, off-diagonal, or complex structure moduli, the dilaton $S$, scalars 
charged under an anomalous U$(1)$ group \cite{Giedt:2000es}, or even discrete torsion 
\cite{Vafa:1986wx}. It is unlikely that all of these could contribute simultaneously, 
therefore it seems reasonable to take each possibility in turn and study it in detail, 
in order either to rule it out or, in the case of positive results, to motivate attempts
to embed it in a more complete model. We choose the diagonal $T$ moduli since their dynamics
and effect on low-energy physics are the best known and do not require excessively 
complicated calculation. Note that the dilaton may also contribute to CP violation in the soft 
terms, but Yukawa couplings, which occur in the superpotential, cannot have $S$-dependence;
then the dilaton cannot alone be the source of CP violation (see also \cite{Dent01}). The 
dynamics of $S$ are such that the contribution to soft terms either vanishes, or it depends on
nonperturbative corrections which are currently uncalculable (see sec.~\ref{sec:further}); 
therefore we leave open the possibility that this contribution is nonzero in the
relevant case (sec.~\ref{sec:softcalc}), but no definite statement is made on this 
point. 

The effective theory is believed to be invariant under target-space modular 
transformations \cite{FerraraLST,FerraraLT} for which the modulus $T^i$ transforms as 
\begin{equation}
	T^i \rightarrow \frac{\alpha T^i-i\beta}{i\gamma T^i+\delta}
\end{equation}
where $\alpha$, $\beta$, $\gamma$, $\delta$ are integers satisfying 
$\alpha\delta-\beta\gamma = 1$ \cite{DijkgraafVV}. The moduli $T_i$, $i=1,2,3$ are defined 
as $iT^i=2(B_{2i-1,2i}+i\sqrt{\det G^i})$, where $i$ labels the three complex planes of the
orbifold, $B_{mn}$, $G_{mn}$ are the antisymmetric tensor field and metric respectively 
in the compact directions and $G^i$ is the $i$'th 2-by-2 submatrix of $G_{mn}$ (assumed block
diagonal). We simplify to the case of a single overall modulus $T^i=T$. The dilaton $S$ 
\cite{DerendingerFKZ} and matter fields \cite{LauerMN91} may also have nontrivial
transformation properties. This symmetry has many implications for the behaviour of the
effective theory: no observable modulus-dependent quantity can change if the v.e.v.\ of 
$T$ is replaced by its image under transformation. In addition the underlying theory 
is exactly CP invariant, since CP is a discrete gauge symmetry of strings 
\cite{ChoiKN,DineLM}, thus $\vev{T}$ and $\vev{T^*}$ describe formally equivalent vacua. 

Then the values at which $\vev{T}$ is stabilized are important in finding the 
modulus-dependent low-energy couplings, in particular the Yukawa couplings and the scalar 
soft SUSY-breaking terms. Below, we parameterize the unknown dilaton dynamics by assuming that
the dilaton v.e.v.\ is constant (up to a well-defined loop correction that is required to 
preserve modular invariance) and varying the contribution of the dilaton to the potential
$V_{\rm dil}\propto |W^S/W + K^S|^2$ within reasonable 
limits. This is equivalent to assuming that the stabilization mechanism for the $T$-modulus 
has essentially no effect on the the value of the dilaton, and thus that any variation of 
the dilaton v.e.v.\ and $F$-term can be neglected. This assumption has since been tested in 
\cite{KhalilLM}, where essentially the same range of models for the dynamics of $T$ was 
considered, but the effects of different mechanisms for dilaton stabilization were worked 
out in detail. The results are striking: in most cases the dilaton v.e.v.\ is indeed 
constant, despite some variation in the values of $\vev{T}$ obtained. Then our previous 
results, and those of \cite{Bailin:1998xx,Bailin:1998iz}, remain useful, and in fact 
concide with some of the phenomenologically interesting minima found in \cite{KhalilLM}. 
There are however some discrepancies, which may be due in part to nontrivial variation of the 
dilaton away from the global minimum; but for the mechanisms which determine $\vev{S}$ 
uniquely we still find minima in $T$ that do not occur in \cite{KhalilLM}.

It was shown in \cite{Dent01} that in orbifold models with three matter generations, CP 
violation in Yukawa couplings cannot originate from a modulus v.e.v.\ on the boundary of 
the fundamental domain ${\mathcal F}$, in other words satisfying either $|\vev{T}|=1$ or 
Im$\,\vev{T}=1/2+n$, $n$ integer. This resulted from an interplay between the modular and 
CP invariances. Since in many cases $T$ is stabilized at precisely these values 
\cite{CveticFILQ}, this seems to rule out many scenarios for the origin of CP violation. 
Recently, however, a nonzero value of the Jarlskog parameter $J_{CP}$, which derives from 
the Standard Model Yukawa couplings \cite{Jarlskog85} was claimed to occur for $T$ on the 
unit circle \cite{Lebedev01}, in a model with more than three generations where some fields 
are assumed to become heavy by an unspecified mechanism. The previous result is evaded by 
taking the three light generations to mix with the heavy matter under a modular transformation. 
The calculated form of $J_{CP}(T)$ is not modular invariant, which appears suspicious from 
the point of view of an exactly invariant theory. In fact for $|T|=1$, $J$ is sent to $-J$ 
under the duality $T\rightarrow 1/T$. Minima on the unit circle are generic, so 
it is important to understand this result and investigate whether it is consistent with an 
explicit mechanism for decoupling the unwanted matter. This point will be addressed in a 
future publication.

\section{Soft supersymmetry-breaking terms} \label{sec:soft_withU}
We calculate soft supersymmetry-breaking terms explicitly for particular models 
motivated by perturbative heterotic string theory, and find the dependence on the dilaton 
and moduli. This will enable us to look in more detail at the scenario in which CP violating
phases appear in the scalar trilinear $A$-terms and bilinear $B$ term due to a complex 
v.e.v.\ of the $T$ modulus. We will concentrate on the modular transformation properties of 
the soft terms, and the question of whether deviations from the truncated value of the 
condensate induce significant changes in the calculation of soft terms. 

\subsection{Calculation of soft terms} \label{sec:softcalc}
By coupling the visible sector, which has just the field content of the MSSM, to 
supergravity and a hidden sector in which supersymmetry is broken, the low-energy theory
appropriate for calculating phenomenology can be found \cite{SoniWeldon}. The procedure is 
described in more 
detail in, for example, \cite{BrignoleIMx2,KaplL93}. In principle one should find the 
vacuum of the full theory by minimising the scalar potential in terms of all the fields,
including visible and hidden matter; in practice for a string scale near the Planck scale 
and a hidden sector with a strong interaction scale a few orders of magnitude below, the 
effects of the visible sector v.e.v.'s will be negligible.

To define the full theory, the superpotential and K\"ahler potentials for the visible 
and hidden sectors are simply added. The hidden sector superpotential takes the 
Veneziano-Yankielowicz form \cite{VenezianoY} 
\be
W_{\rm np} = \frac{f_{\rm g}(S,T)}{4}U - \frac{b_0}{96\pi^2}U \ln(kU)
\ee
where $b_0\equiv -3c(G)$ is the one loop beta-function coefficient for a SYM hidden 
sector without matter. The gaugino condensate is described by the classical, 
composite field $U\equiv \vev{W^{\alpha a}W_\alpha^a}/S_0^3$, where $S_0$ is the conformal 
compensator in the superconformal formulation of matter-coupled supergravity 
\cite{BurgessDQQ,SUGRA_matter}. The hidden sector gauge kinetic function 
$f_{\rm g}=S+\Delta(T)/(16\pi^2)$ includes a modulus-dependent threshold correction 
$\Delta(T)$ \cite{DixonKL91}. On ``integrating out'' the condensate via 
$\partial W/\partial U=0$ we find the standard ``truncated superpotential'' 
\be
W^{\rm (tr)}(S,T)=\frac{b_0}{96\pi^2}U^{\rm (tr)}
\ee
where
\be 
U^{\rm (tr)} = e^{24\pi^2 S/b_0 - 1} (kh(T))^{-1}, \label{eq:Utr}
\ee
and $h(T)$ is defined by $h(T)\equiv e^{3\Delta(T)/2b_0}$.
For the theory to be invariant under SL$(2,{\mathbb Z})$ modular transformations acting
on $T$ we must have the form 
$h(T)= \eta^{6-18\delta_{\rm GS}/b_0}(T) H(T)^{-1}$, where $\eta(T)$ is the Dedekind eta 
function, $\delta_{\rm GS}$ is a numerical coefficient associated with the cancellation of 
modular anomalies \cite{DerendingerFKZ} and $H(T)$ is a modular invariant function without 
singularity in the fundamental domain \cite{CveticFILQ}. 

For the visible sector we take
	\[ W_v = \mu_{Wij}\Phi_i\Phi_j + y_{ijk}\Phi_i\Phi_j\Phi_k \]
where $\Phi_i$ are the MSSM chiral matter superfields and $y_{ijk}$, $\mu_{Wij}$ are
respectively the coefficients of Yukawa couplings, and a possible supersymmetric bilinear
coupling which is required to couple the Higgs doublets of the MSSM in order to obtain 
correct electroweak symmetry-breaking (the ``mu-term''). For this to occur, the parameter 
$\mu$, which has the dimensions of a mass, must be of the same order as the soft 
supersymmetry-breaking masses. This seems to require a fine-tuning of the underlying
theory, since dimensionful parameters in $W$ are naturally of the order of the fundamental
scale $M_P$. The ``mu problem'' in supersymmetric phenomenology is the question of how the 
hierarchy between $M_P$ and $\mu$ can be generated. Various solutions have been proposed,
notably a non-minimal mixing of the two Higgs doublets in the K\"ahler potential 
\cite{GiudiceM}, which has an effect analogous to the supersymmetric $\mu_W$ term,
generating a Higgsino mass term and a scalar bilinear $B\mu$ term after 
supersymmetry-breaking. Such a term in the K\"ahler potential can appear naturally if the 
Higgs superfields are in the untwisted sector of an orbifold compactification 
\cite{AntoniadisGNT_mu}, however the resulting couplings cannot easily be written down in a 
form which is manifestly modular invariant (see \cite{KhalilLM} for a recent discussion).
For simplicity we use the alternative option, that the term $\mu H_U H_D$ is 
present in the superpotential at the string scale and is tuned to be of the right order 
of magnitude by some unspecified mechanism. In this case we allow for modulus-dependent 
couplings $\mu_{ij}(T)$, $y_{ijk}(T)$, which are in general required to maintain modular 
invariance.

If we assume that the matter fields are in the twisted sector of an orbifold
compactification, the K\"ahler potential is
	\[ K_v = \sum_i(T+T^\dag)^{n_i}\Phi_i^\dag\Phi_i \]
to second order in $\Phi_i$, where $n_i$ is the modular weight of $\Phi_i$. 
The dilaton and modulus K{\" a}hler potential $\hat{K}(S,T)$ is taken to be of the form
\begin{equation}
	\hat{K} = P(y) -3 \ln(T+T^\dag) 
\end{equation}
where $y=S+S^\dag -1/(8\pi^2)\cdot 3\delta_{\rm GS}\ln(T+T^\dag)$ is a modular invariant 
combination of fields such that $g^{-2}_{\rm string}={\rm Re}\,y$. The perturbative 
string K\"ahler potential for the dilaton $K_{\rm pert}=-\ln(y)$ has been replaced by a 
real function $P(y)$ which 
parameterizes stringy nonperturbative dynamics, which are hoped to contribute to 
dilaton stabilization \cite{Shenker90}, \cite{BanksDine+BGW+ChoiKK,Casas96}. 
We initially take the Green-Schwarz coefficients $\delta_{\rm GS}$ to be zero to simplify the 
calculations, but when calculating the scalar potential for the modulus we relax this
assumption. The correct form of $P(y)$ is not known, however it is possible to constrain 
it by looking for a stable minimum in the potential for the dilaton and requiring 
$P''(y)>0$ to obtain the right sign kinetic term. The complete hidden sector K\"ahler 
potential is then taken to be 
\begin{equation}
	K_{\rm h}(U,S,T) = \hat{K} - 3\ln\left(1- Ae^{\hat{K}/3} (UU^\dag)^{1/3}\right) 
	\label{eq:K1} 
\end{equation} 
where $A$ is a constant. It was shown in \cite{BurgessDQQ} that this expression for
the K\"ahler potential of $U$ has the correct dependence on $S$ and $T$, as well as
being modular invariant. However, it can only be determined up to a constant factor,
and may receive higher-order corrections. The constant $A$ cannot at present be
computed, due to our incomplete knowledge of supersymmetric gauge dynamics, but is
expected to be of order unity.

The trilinear $A$-terms and the scalar bilinear $B$ term corresponding to $y_{ijk}$ and 
$\mu_{ij}$ respectively can be extracted from the standard formula for the scalar 
potential 
\begin{equation}
	V = e^{K}\left((W^*_I+K_IW^*)(K^{-1})_J^I
	(W^J+K^JW) - 3|W|^2\right) \label{eq:V_KW}
\end{equation}
where the indices run over $U,S,T$ and $\phi_i$ ($\phi_i$ being
the scalar component of $\Phi_i$) and we assume a single hidden sector gauge group. We
find a general formula for soft SUSY-breaking scalar interactions: 
\begin{eqnarray*}
	\delta{\mathcal L}_{(scalar)}&=&-\frac{b_0z^{*3}e^{\hat{K}/2}}{96\pi^2(1-A|z|^2)^2}
	\left\{ W_v\cdot \left[ \frac{P'(y)}{P''(y)}\left(P'(y)-\frac{\omega'(S^*)}
	{\omega(S^*)}\right) + (T+T^*) \frac{h'(T^*)}{h(T^*)} \right]\right. \\
	&+& \sum_i\phi_i \frac{\partial W_v}{\partial \phi_i}\cdot 
	\left[n_i\left(1+\frac{1}{3}(T+T^*)\frac{h'(T^*)}{h(T^*)}\right) - 
	\frac{\ln(e^{-\hat{K}/2}z^{*3}\omega(S^*)h(T^*))}{1-A|z|^2} \right] \\
	&+& \left. \frac{\partial W_v}{\partial T}\cdot 
	-(T+T^*)\left(1+\frac{1}{3}(T+T^*)\frac{h'(T^*)}{h(T^*)} \right) \right\} + h.c.
\end{eqnarray*}
where we include terms up to third order in the visible sector fields and to all orders 
in the dilaton, modulus and hidden sector condensate $z$ defined by $z^3=e^{\hat{K}/2}U$, 
taking $\delta_{\rm GS}=0$. The function $\omega(S)\equiv e^{-24\pi^2S/b_0}$ is defined 
analogously to $h(T)$.

If the visible sector superpotential contains a coupling $y_{123}(T)\Phi_1\Phi_2\Phi_3$ 
then the Lagrangian will contain the trilinear interaction
\begin{eqnarray} 
	A_{123}y_{123}\phi_1\phi_2\phi_3 &=& -y_{123}\phi_1\phi_2\phi_3\frac{b_0}{96\pi^2} 
	\frac{e^{\hat{K}/2}{z^*}^3}{(1-A|z|^2)^2} \left[ \frac{P'(y)}{P''(y)} \left(P'(y) 
	-\frac{\omega'(S^*)}{\omega(S^*)}\right) \right. \nonumber \\
	&+& \left(3+ (T+T^*) \frac{h'(T^*)}{h(T^*)} \right) \left(1+ \sum_{i=1}^3 
	\frac{n_i}{3} -\frac{(T+T^*)}{3}\frac{y'_{123}(T)}{y_{123}(T)} \right) \nonumber \\
	&+& \left. 3\left(-\frac{\ln(e^{-\hat{K}/2}z^{*3}\omega(S^*)h(T^*))}{1-A|z|^2} -1 
	\right) \right]. \label{eq:AwU}
\end{eqnarray}
The auxiliary fields $F^S$ and $F^T$ as defined by
\begin{equation}
	F^i = e^{K/2}(K^{-1})^i_j(W^j+K^jW). \label{eq:F_SUGRA}
\end{equation}
are
	\[ F^S = \frac{b_0}{96\pi^2} |z|^3 (1-A|z|^2)^{-1/2} \frac{1}{P''(y)} \left(P'(y)- 
\frac{\omega'(S)}{\omega(S)}\right), \]
	\[ F^T = - \frac{b_0}{96\pi^2} |z|^3 (1-A|z|^2)^{-1/2} \frac{(T+T^*)}{3} 
\left(3+ (T+T^*) \frac{h'(T)}{h(T)} \right). \]
Superscripts denote holomorphic indices, subscripts antiholomorphic ones, so $F_T=(F^T)^*$.
The A-term can then be written as
	\[ A_{123} = -\left(\frac{z^*}{|z|}\right)^3 e^{K/2} \left[\hat{K}^SF_S + \hat{K}^T
	F_T \left(1+\sum_{i=1}^3 \frac{n_i}{3} - \frac{T+T^*}{3}\frac{y'_{123}(T)}
	{y_{123}(T)} \right) \right. \]
	\[ \left. +3\frac{|b_0|}{96\pi^2}\frac{|z|^3}{(1-|z|^2)^{1/2}} \left(-\frac{
	\ln(e^{-\hat{K}/2}z^{*3}\omega(S^*)h(T^*))}{1-A|z|^2} -1 \right) \right]. \] 
Similarly, for a coupling $\mu_{12}(T)\Phi_1\Phi_2$ in $W_v$ a corresponding soft 
supersymmetry-breaking coupling $B\mu_{12}\phi_1\phi_2$ is generated, with
\begin{eqnarray}
	B &=& -\frac{b_0}{96\pi^2}\frac{e^{\hat{K}/2}{z^*}^3}{(1-A|z|^2)^2}
	\left[ -1 + \frac{P'(y)}{P''(y)} \left(P'(y) 
	-\frac{\omega'(S^*)}{\omega(S^*)}\right) \right. \nonumber \\
	&+& \left(3+ (T+T^*) \frac{h'(T^*)}{h(T^*)} \right) \left(1+ \sum_{i=1}^2 
	\frac{n_i}{3} -\frac{(T+T^*)}{3}\frac{\mu'_{12}(T)}{\mu_{12}(T)} \right)\nonumber \\
	&+& \left. 2\left(-\frac{\ln(e^{-\hat{K}/2}z^{*3}\omega(S^*)h(T^*))}{1-A|z|^2} -1 
	\right) \right]. \label{eq:BwU}
\end{eqnarray}
In these expressions the truncated approximation for $z$, equivalent to imposing 
$U=U^{\rm (tr)}$ (Eq.~(\ref{eq:Utr})), is implemented by neglecting $A|z|^2$ next to 1 
and setting the logarithm in the last term equal to $(-1)$; we can also use the formula 
for the gravitino mass $m_{3/2}\equiv e^{K/2}|W|=|b_0z_{\rm tr}^3|/(96\pi^2)$ to simplify 
the prefactors and make contact with previous results. Deviations from the truncated 
approximation are treated in detail in \cite{me99} (see also \cite{Goldberg95}).

\subsection{Phenomenological discussion}\label{sec:softphenom}
The complex phases and flavour structure of the soft breaking terms are mainly determined by
the dependence on the $T$ modulus: the $F_S$ terms are universal and, for a single 
condensate, real. Apart from the phase of $z^{*3}$ (which is eliminated by the redefinition 
of fields in going to the softly broken globally supersymmetric theory, for which see below) 
a complex phase can only enter through the auxiliary field $F^T$ and the
modulus-dependent couplings $y(T)$, $\mu(T)$. 
When $T$ is stabilized at a minimum of the scalar potential (section \ref{sec:V_T}), 
$F^T$ may be zero, real or have a complex phase which is of order $0.1$--$1$. If the
term involving the derivative of $y_{123}(T)$ were absent, the $A$-terms would have a
common phase, that of $F^T$, and their magnitudes would be determined by the modular
weights $n_i$. Using an explicit formula for Yukawa couplings, Khalil {\em et al.}\/\
\cite{KhalilLM} recently showed that these logarithmic derivatives were real to a good 
approximation in some cases of interest, thus a common phase for $A$-terms may be a good 
approximation.

In the special case of all $n_i$ equal we would recover the ``minimal supergravity'' 
ansatz for the soft terms, in which the trilinear couplings are proportional to $y_{ijk}$, 
{\em i.e.}\/\ $A_{ijk}=A$ for all $i,j,k$. However in general the $A$-terms will be 
non-universal, due to the different values of $n_i$ and the terms involving 
$y'_{ijk}/y_{ijk}$ (which are also essential for modular invariance): their magnitudes and 
phases will be different and there will be {\em off-diagonal}\/ (and complex) $A$-terms in 
the super-KM basis. Similarly the $B\mu$ bilinear coupling may have a phase different from 
that of $\mu$ (only the phase difference between $B\mu$ and $\mu$ is physically observable) 
which will feed through into a complex mass matrix for charginos and neutralinos at low 
energies. This is a phenomenologically interesting scenario, which may result in predictions 
for CP-violating observables which differ significantly from the SM. However it is
severely restricted by the non-universality of scalar masses which would result if the
matter generations of the MSSM had different modular weights: any departure from
degeneracy of scalar masses is likely to result in contributions to flavour changing
neutral current processes in excess of the experimental limits (see {\em e.g.}\/\ 
\cite{Gabbiani:1996hi}).

In the light of this discussion, deviations from the truncated approximation in the formula
(\ref{eq:AwU}) do not have an important direct effect. New complex phases are not 
introduced and the corrections arising from $z\neq z_{\rm tr}$ are universal, {\em i.e.}\/\
flavour-independent. However the overall magnitude of the soft terms may be slightly 
changed, a ``second-order'' effect. So from now on we will use the standard formulas 
resulting from the truncated approximation:
\begin{equation}
	A_{123} = -\left(\frac{z^*}{|z|}\right)^3 e^{\hat{K}/2} \left[\hat{K}^SF_S + 
	\hat{K}^TF_T \left(1+\sum_{i=1}^3 \frac{n_i}{3} - 
	\frac{T+T^*}{3}\frac{y'_{123}(T)}{y_{123}(T)} \right) \right] \label{eq:Atr}
\end{equation}
and
\begin{equation}
	B = \left(\frac{z^*}{|z|}\right)^3 e^{\hat{K}/2} \left[-m_{3/2} -\hat{K}^SF_S -
	\hat{K}^TF_T \left(1+\sum_{i=1}^2 \frac{n_i}{3} - \frac{T+T^*}{3}
	\frac{\mu'_{12}(T)}{\mu_{12}(T)} \right) \right] \label{eq:Btr}
\end{equation}
where the auxiliary fields are now
	\[ F^S = -m_{3/2} \frac{1}{P''} \left(P'- \frac{\omega'(S)}{\omega(S)}\right), \]
\begin{equation}
	F^T = m_{3/2} \frac{(T+T^*)}{3} \left(3+ (T+T^*) \frac{h'(T)}{h(T)} \right). 
	\label{eq:FSTtr}
\end{equation}
The corresponding formulae in the case $\delta_{\rm GS}\neq 0$, which will be needed when 
discussing the effect of changing $\delta_{\rm GS}$ on the minimisation of $T$ and on the soft 
terms, have been derived in \cite{deCarlos:1993pd,Bailin:1997fh}.
As is well known, not all complex couplings in the Lagrangian result in 
CP violation. We must consider whether the phases are physical and how many can be 
eliminated by redefinition of fields. In particular, the behaviour of the $A$-terms under 
SL$(2,{\mathbb Z})$ modular transformations must be found, since physical quantities should 
be modular invariant and (in general) the individual couplings will not be. Little can be 
deduced from the phase of a single $A$-term without considering the whole set of couplings. 
The results (\ref{eq:AwU}, \ref{eq:Atr}) have nontrivial modular transformations, 
particularly when the observable matter fields are mixed by modular transformations. 
Since we suppose only a single $\mu$-term in the superpotential, modular transformations
cannot mix the bilinear $B\mu$ coupling with any other and it is relatively easy to verify 
that this term is modular invariant. So we will focus on the trilinear $A$-terms.

\subsection{Modular transformations of soft terms} \label{sec:mod_soft}
Recall that the superpotential is a modular form of weight $-3$, so under modular 
transformations
	\[ U \mapsto \frac{\zeta_W U}{(i\gamma T+\delta)^3},\qquad 
	W_v \mapsto \frac{\zeta_W W_v}{(i\gamma T+\delta)^3}. \]
where $\zeta_W$ is a phase depending only on the parameters $\alpha$, $\beta$, $\gamma$, 
$\delta$ of the transformation. The observable fields transform into one another under 
SL$(2,{\mathbb Z})$ as
	\[ \Phi_i \mapsto C_{im} (i\gamma T+\delta)^{n_i} \Phi_m \]
where $C_{im}$ is a unitary matrix and $n_i=n_m$ for all $i,m$ such that $C_{im}\neq 0$,
thus the Yukawa couplings are constrained to transform as
	\[ y_{ijk}(T) \mapsto(i\gamma T+\delta)^{(-\sum n_i -3)} \tilde{y}_{ijk}(T) \]
where
	\[ \tilde{y}_{ijk} C_{im}C_{jn}C_{kp} = \zeta_W y_{mnp} \]
and $\sum n_i$ denotes the sum over the three indices, {\em i.e.}\/\ $(n_i+n_j+n_k)$. This 
implies that
	\[ \tilde{y}_{ijk}(T) = \zeta_W y_{mnp}(T)
	C^\dagger_{mi}C^\dagger_{nj}C^\dagger_{pk}. \]
We rewrite the scalar trilinear interactions as
	\[ A_{ijk}y_{ijk} \phi_i\phi_j\phi_k = -e^{\tilde{K}/2} \frac{{z^*}^3}{|z|^3}\phi_i\phi_j\phi_k \]
	\[ \cdot \left[\hat{K}^SF_S 
	y_{ijk} + \hat{K}^TF_T \left(\left(1+ \frac{\sum n_i}{3}\right) y_{ijk}(T) - 
	\frac{(T+T^*)}{3} y'_{ijk}(T) \right)\right]. \]
The quantity $\hat{K}^TF_T = b_0/(96\pi^2)|z_{\rm tr}|^3 (3+ (T+T^*)h'(T)/h(T))$ transforms as
	\[ \hat{K}^TF_T \mapsto\frac{(-i\gamma T^*+\delta)^2}{|i\gamma T+\delta|^2} 
	\hat{K}^TF_T \]
where we have used the fact that $h(T)$ must have modular weight $3$. Also, under 
modular transformations the expression 
	\[ (1+\sum n_i/3)y_{ijk} -\frac{1}{3}(T+T^*)y'_{ijk}(T) \] 
goes to 
	\[ \zeta_W C^\dagger_{mi}C^\dagger_{nj}C^\dagger_{pk} 
	(i\gamma T+\delta)^{-3-\sum n_i}\frac{(i\gamma T+\delta)^2}{|i\gamma T+\delta|^2}
	\left(\left(1+ \frac{\sum n_m}{3}\right) y_{mnp}(T) - 
	\frac{(T+T^*)}{3} y'_{mnp}(T) \right) \]
since the matrices $C_{ij}$ and the phase $\zeta_W$ do not depend on $T$. 
Recalling that $z$ is defined as $(e^{\tilde{K}/2}U)^{1/3}$, we have
	\[ \frac{{z^*}^3}{|z|^3} \mapsto \zeta_W^* \left(\frac{-i\gamma T 
	+\delta}{|i\gamma T+\delta|}\right)^{-3} \frac{{z^*}^3}{|z|^3} \]
under modular transformations. It is now easy to confirm that the expression 
$A_{ijk}y_{ijk} \phi_i\phi_j\phi_k$ is modular invariant, since $C^\dagger_{mi}C_{ij} 
=\delta_{mj}$ and all phases and factors of $(i\gamma T+\delta)$ cancel.

This also implies that $A_{ijk}y_{ijk}$ (no sum!) transforms ``inversely'' to 
$\phi_i\phi_j\phi_k$:
	\[ (Ay)_{ijk} \mapsto (i\gamma T+\delta)^{-\sum n_i} (Ay)_{mnp} 
	C^\dagger_{mi}C^\dagger_{nj}C^\dagger_{pk}. \] 
To clarify notation here we write $(Ay)_{ijk}\equiv A_{ijk}y_{ijk}$, where no sum is 
implied on the RHS. Then dividing by $y_{ijk}$ and its modular transform we find 
\begin{equation} 
	A_{ijk} \mapsto (i\gamma T+\delta)^3\zeta_W^{-1} \frac{\sum_{mnp}(Ay)_{mnp} 
	C^\dagger_{mi}C^\dagger_{nj}C^\dagger_{pk}}
	{\sum_{m'n'p'}y_{m'n'p'}C^\dagger_{m'i}C^\dagger_{n'j}C^\dagger_{p'k}}. 
	\label{eq:Aijk_transf}
\end{equation}
If we ignore the unitary mixings $C_{ij}$ by assuming $C_{ij}=e^{i \theta_i}\delta_{ij}$ 
then all the phases $\theta_i$ cancel and the $A$-terms transform with a
universal factor $(i\gamma T+\delta)^3\zeta_W^{-1}$.

\subsection{Rescaling to a global SUSY with soft breaking terms} \label{sec:rescaling} 
The formulas presented so far have been in terms of fields normalised as in the effective 
supergravity theory. It is usual to rescale the visible sector component fields and rotate 
away the phase of the hidden sector superpotential, so that the low-energy theory is just 
the MSSM with {\em canonical}\/ kinetic terms for the chiral matter, and soft breaking 
terms expressed in terms of the gravitino mass $m_{3/2}=|b_0/(96\pi^2)z^3|$ 
\cite{SoniWeldon}. The scalar potential is then written in terms of normalised fields 
$\hat{\phi}_i = K_i^{1/2} \phi_i$, where $K_i(T) = (T+T^*)^{n_i}$. The Yukawa couplings 
are rescaled as 
\begin{equation}
	\hat{y}_{ijk} = \frac{W^*}{|W|} e^{\hat{K}/2} 
	(K_iK_jK_k)^{-1/2}y_{ijk} 
	\label{eq:yhat_ijk}
\end{equation}
so that 
	\[ \hat{y}_{ijk} \hat{\phi}_i\hat{\phi}_j\hat{\phi}_k = 
	\frac{W^*}{|W|} e^{\hat{K}/2} y_{ijk} \phi_i\phi_j\phi_k \]
which is modular invariant. We have $W^*/|W| = (z^*/|z|)^3$ {\em in vacuo}\/, and since
	\[ A^{\rm SUGRA}_{ijk} y_{ijk} \phi_i\phi_j\phi_k = A^{\rm SUSY}_{ijk} 
	\hat{y}_{ijk} \hat{\phi}_i\hat{\phi}_j\hat{\phi}_k \]
we deduce that
\begin{eqnarray*}
	A^{\rm SUSY}_{ijk} &=& \frac{|z|^3}{{z^*}^3} e^{-K/2} A^{\rm SUGRA}_{ijk} \\
	&=& -\hat{K}^SF_S -\hat{K}^TF_T \left(1+ \frac{n_i+n_j+n_k}{3} - \frac{(T+T^*)}{3} 
	\frac{\partial \ln y_{ijk}}{\partial T} \right) 
\end{eqnarray*}
\begin{equation} 
	= m_{3/2} \left[\frac{P'}{P''} \left(P'-\frac{\omega'(S^*)}{\omega(S^*)}\right) 
	+ \left(3+ (T+T^*) \frac{h'(T^*)}{h(T^*)} \right) \left(1+ \frac{\sum n_i}{3} - 
	\frac{(T+T^*)}{3}\frac{y'_{ijk}(T)}{y_{ijk}(T)} \right)\right] 
\end{equation}
has modular weight zero. In particular, the transformation of the factors $W^*/|W|$ and 
$e^{\hat{K}/2}$ cancels the previously-noted factor of
$(i\gamma T+\delta)^3\zeta_W^{-1}$, so the expression is exactly invariant if unitary 
mixings are neglected. 

However, the $A^{\rm SUSY}$ still in general transform with the complicated 
mixing in terms of the $y_{ijk}$ and $C_{im}$ as indicated in (\ref{eq:Aijk_transf}). 
These involved transformation properties are a major obstacle to finding the implications 
of complex phases in the $A$-terms for physical CP-violating quantities. For a particular 
v.e.v.\ of $T$ there will be certain predictions for the complex phases of the $A$-terms, 
but in general these predictions will not be invariant under the modular transformation 
acting on $T$ and on the matter fields. We expect that physical quantities should depend on 
combinations of the couplings in ${\mathcal L}$ that are modular invariant, however in the
absence of a realistic model, and in the general case where ``modular eigenstates'' cannot 
be defined, it is not clear how to construct the relevant quantities. 

\section{Stabilizing string moduli}\label{sec:V_T}
\subsection{Introduction}
The breaking of local supersymmetry leads to a non-vanishing scalar potential for the 
``flat directions'' of string theory, the dilaton and compactification moduli. This 
suggests the possibility that these quantities are dynamically determined after 
supersymmetry is broken. When supersymmetry-breaking is mediated by gravity the scalar 
potential $V(S,T)$ is of order $|F|^2$, where $F$ denotes a super\-symmetry-breaking 
auxiliary field, while the values of the dilaton and moduli fields
vary over the Planck scale (when the units of $S$ and $T$ are restored) so the flat 
directions acquire masses of order $m_{3/2}\sim\,$TeV. While this scenario is not without 
drawbacks for cosmology, and in most models predicts a large negative cosmological constant
at the minimum of $V$, it appears more promising for phenomenology, since the 
dynamics of moduli can be studied through an effective field theory and it is possible in
principle to make predictions based on specific super\-symmetry-breaking mechanisms.

In the models of super\-symmetry-breaking via gaugino condensation that we have considered,
the dependence of the scalar potential on the dilaton and moduli is determined by the gauge
kinetic function of the hidden sector gauge group and the K\"ahler potential for $S$ and
$T$. We consider the simplest hidden sector consisting of a single gauge group factor 
without matter, however we allow the Green-Schwarz coefficient to be non-zero. For 
convenience we quote the scalar potential from \cite{Taylor90,deCarlosCM91,me99} 
({\em without}\/ assuming the truncated approximation):
\begin{equation} 
	V = \left(\frac{b_0}{96\pi^2}\right)^2 \frac{|z|^4}{(1-A|z|^2)^2} 
	\left\{\frac{3}{A}\left| 1+\ln\left(\omega(S)h_1(T) e^{-\hat{K}/2} z^3\right)
	\right|^2 + C_2'(S,T) |z|^2 \right\} 
	\label{eq:VSTZ_1again}
\end{equation}
where 
	\[ z= (e^{\hat{K}/2}U)^{1/3} = e^{P(y)/6} (T+T^*)^{-1/2} U^{1/3} \]
as before, and
\begin{eqnarray}
	C_2'(S,T) &=& \left(\frac{1}{P''(y)} \left|P'(y)- \frac{\omega'(S)}{\omega(S)} 
	\right|^2 + \right. \nonumber \\
	&+& \left.\frac{1}{3(1+\frac{\delta_{\rm GS}}{8\pi^2}P'(y))} 
	\left|3\left(1-\frac{3\delta_{\rm GS}}{b}\right) + (T+T^*)\frac{h_1'(T)}{h_1(T)} 
	\right|^2 -3 \right). \nonumber \\ & &
	\label{eq:C2'}
\end{eqnarray}
Recall that $\omega(S)=e^{-24\pi^2S/b_0}$, and we take the ansatz 
	\[ h_1(T) = \eta^{6-18\delta_{\rm GS}/b_0}(T)\left(H(T)\right)^{-1} \]
where $H(T)$ is in general a modular invariant function without singularities in the
fundamental domain. This form for $h_1$ is supposed to originate from the threshold correction 
	\[ \Delta_a(T) = -(b_0-3\delta_{\rm GS}) \ln\left(|\eta^4(T)| (T+T^*)\right) 
	+ (b_0/3) \ln |H(T)|^2, \] 
with $H$ a holomorphic function of $T$.

Note that modular invariant, so-called ``universal'' threshold corrections in heterotic string 
theory have been calculated \cite{Kiritsis:1997dn,Nilles:1997vk} which take the form 
\[ \Delta_a = -(b_0-3\delta_{\rm GS}) \ln\left(|\eta^4(T)| (T+T^*)\right) - k_a Y(T) \] 
where $Y(T)$ is modular invariant but {\em not}\/ the real part of a holomorphic function. As 
with the threshold corrections involving $\eta(T)$, the non-holomorphic part of the threshold 
correction appears in the $T$-dependence of the K\"ahler potential of the effective field 
theory, so the universal threshold corrections imply a correction to both the superpotential 
and K\"ahler potential for the dilaton and modulus. This results in the modular transformation 
properties 
becoming considerably more 
complicated; so these corrections cannot be included in the above prescription for $h_1(T)$. 
Note that the universal corrections $-k_a Y$ can be absorbed by a redefinition of the string 
coupling $g^{-2}_{\rm string}={\rm Re}\,y$, which formally justifies neglecting them 
if they are small. The effect of the universal threshold corrections on stabilizing $T$ has 
been found using the truncated approximation for the gaugino condensate \cite{Bailin:2000ra} 
resulting in a $T$-dependent potential similar in form to $V^{(tr)} \propto 
C'_2|z_{\rm tr}|^6$. We might then anticipate that the full condensate-dependent potential 
would have a similar form to (\ref{eq:VSTZ_1again}) with a different function of $S$ and 
$T$ replacing $C'_2$. However the mathematical complexity of the threshold corrections 
prevented us from investigating further. 

The invariance of the theory under target-space T-duality $T\mapsto 1/T$ will ensure that the 
modulus is stabilised at a value of order unity (assuming that the extra dimensions are 
indeed dynamically compactified, {\em i.e.}\/\ that $V$ becomes large and positive for 
Re$\,(T)$ very large and very small). However, in heterotic string theory no such duality 
applies to the dilaton; without careful model-building either the potential will have no 
minimum in the $S$-direction, or the minimum will lead to an unrealistic value of the gauge 
coupling. This is the well-known dilaton runaway problem, first noticed in the case of a 
single condensing gauge group with the K\"ahler potential for the dilaton taking its string 
tree-level form. There have been various proposals for solving it: the simplest, for our 
purposes, is ``K\"ahler stabilisation'' \cite{Casas96} where the dilaton K\"ahler potential 
is supposed to receive large corrections from nonperturbative string effects \cite{Shenker90}. 
We have assumed that this is the case in deriving the formulas (\ref{eq:VSTZ_1again}) for 
the scalar potential. However, since we are mainly interested in finding the v.e.v.\ of the 
$T$ modulus, the details of the dilaton stabilization will be neglected as far as is 
reasonable. 

\subsection{Calculation procedure}
We will assume that the v.e.v.\ of the dilaton is fixed by nonperturbative effects
irrespective of the value of $T$, such that the unified gauge coupling takes a
phenomenologically reasonable value. The (modular invariant) quantity 
$y=S+S^* - 1/(8\pi^2)\cdot 3\delta_{\rm GS}\ln(T+T^*)$ will be set equal to $4$, which will fix 
the value of $S$ at any given value of $T$. The scalar potential also depends on the 
function $P(y)$ and its derivatives at the point $y=4$. We will treat $P'(y)$ and
$P''(y)^{-1}|P'(y)- \omega'(S)/\omega(S)|^2 \equiv V_{\rm dil}$ as {\em independent} 
parameters and take $P(y)=-\ln 4$, $P'(y)=-1/4$ (the same values as for the perturbative 
K\"ahler potential $K_p=-\ln y$) since the effects of changing these two quantities on the 
potential for $T$ are small. Specifically, changing $P(y)$ would change the overall scale of 
the condensate $z$ but not the shape of the potential, while changing $P'(y)$ would have a
small effect on the prefactor of the second term in (\ref{eq:C2'}). However, 
$V_{\rm dil}$, the first term of $C_2'$, corresponding physically to the amount of
super\-symmetry-breaking originating from the dilaton dynamics, will be treated as a free
(positive semidefinite!) constant parameter. It is proportional to $|F^S|^2$ and may have 
an important effect on the shape of the potential and the cosmological constant.

To determine the shape of the potential as a function of $T$, a number of further 
parameters have to be specified. Apart from $P(y)$ and its derivatives, we require the 
values of $b_0$, $A$, $\delta_{\rm GS}$ and the functional form of $H(T)$ specified by the 
integers $m$ and $n$ and the polynomial $p(J)$ in the expression
	\[ H(T) = (J-1)^{m/2}J^{n/3}p(J). \label{eq:H_Jagain} \]
which is the most general invariant form with no singularities at finite $T$ 
\cite{CveticFILQ}. Of these, $b_0$ and $\delta_{\rm GS}$ can be calculated in specific orbifold 
models for a given hidden sector gauge group; however, we will take the phenomenological 
liberty of varying $b_0$ and $\delta_{\rm GS}$ over typical ranges of values, for the purpose 
of illustration. The constant $A$ is not known and will be set to unity. The form of $H$, 
which parameterizes unknown modular invariant threshold corrections, is essential to 
finding the minimum of the potential. In the absence of definite results from string theory, 
we look at a range of values for $m,n$ and the simplest possibilities for $p(J)$, in an 
approach similar to \cite{CveticFILQ,Bailin:1997fh,Bailin:1998iz}, and look at the possible 
implications for CP violation when $T$ is stabilized at a minimum of the scalar potential. 
In addition, we are able to look at the effects of deviations from the truncated 
approximation, which may be important if they change the position of the minimum in the 
complex $T$ plane. This was not possible in the formalism of
\cite{Bailin:1997fh,Bailin:1998iz}. We calculate the scalar potential in the truncated 
approximation
\begin{equation} 
	V^{\rm (tr)} = \left(\frac{b_0}{96\pi^2}\right)^2 C_2'|z_{\rm tr}|^6 \label{eq:Vtr}
\end{equation}
where $z_{\rm tr}$ and $C_2'$ are functions of $S$ and $T$, and then find the corrections by 
minimising the full scalar potential in the $z$-direction for each value of $T$, with $S$ 
being fixed as described above. Note that the only dependence on $z_{\rm tr}$ is in the 
overall scale of the potential.

\subsection{Results: no dependence on $J(T)$}
The simplest case occurs when the $T$-dependent holomorphic threshold corrections do not
involve the absolute modular invariant $J(T)$, in other words $H={\rm constant}$. This is 
the form that results from a direct perturbative string calculation \cite{DixonKL91}, which 
may however miss universal, modular invariant contributions. The $T$-dependence of the scalar
potential is well-known in this case \cite{FontILQ90,FerraraMTV}, however we are able to 
look in more detail at the effect of changing various parameters.

\begin{figure}[tb]
	\centering
	\includegraphics[width=8.5cm,height=6.5cm]{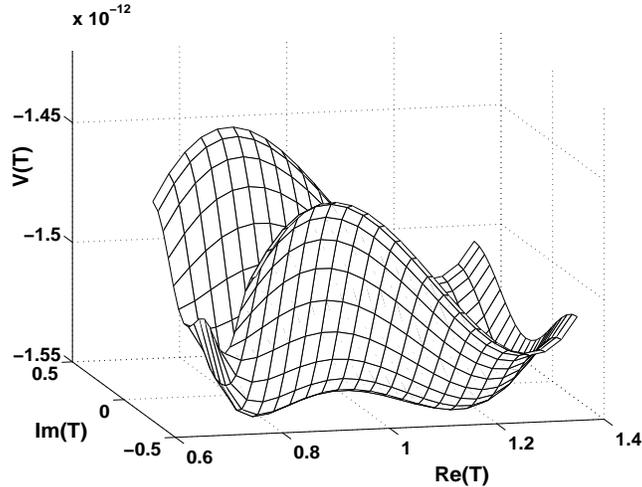}
	\caption{Scalar potential for $m=n=0$, $p(J)=1$, $\beta\equiv b_0/(16\pi^2)=-0.3$, 
	$V_{\rm dil}=0=\delta_{\rm GS}$.}
	\label{fig:VT_FS0d0}
	\end{figure}

\begin{figure}
	\centering
	\includegraphics[width=8.5cm,height=6.5cm]{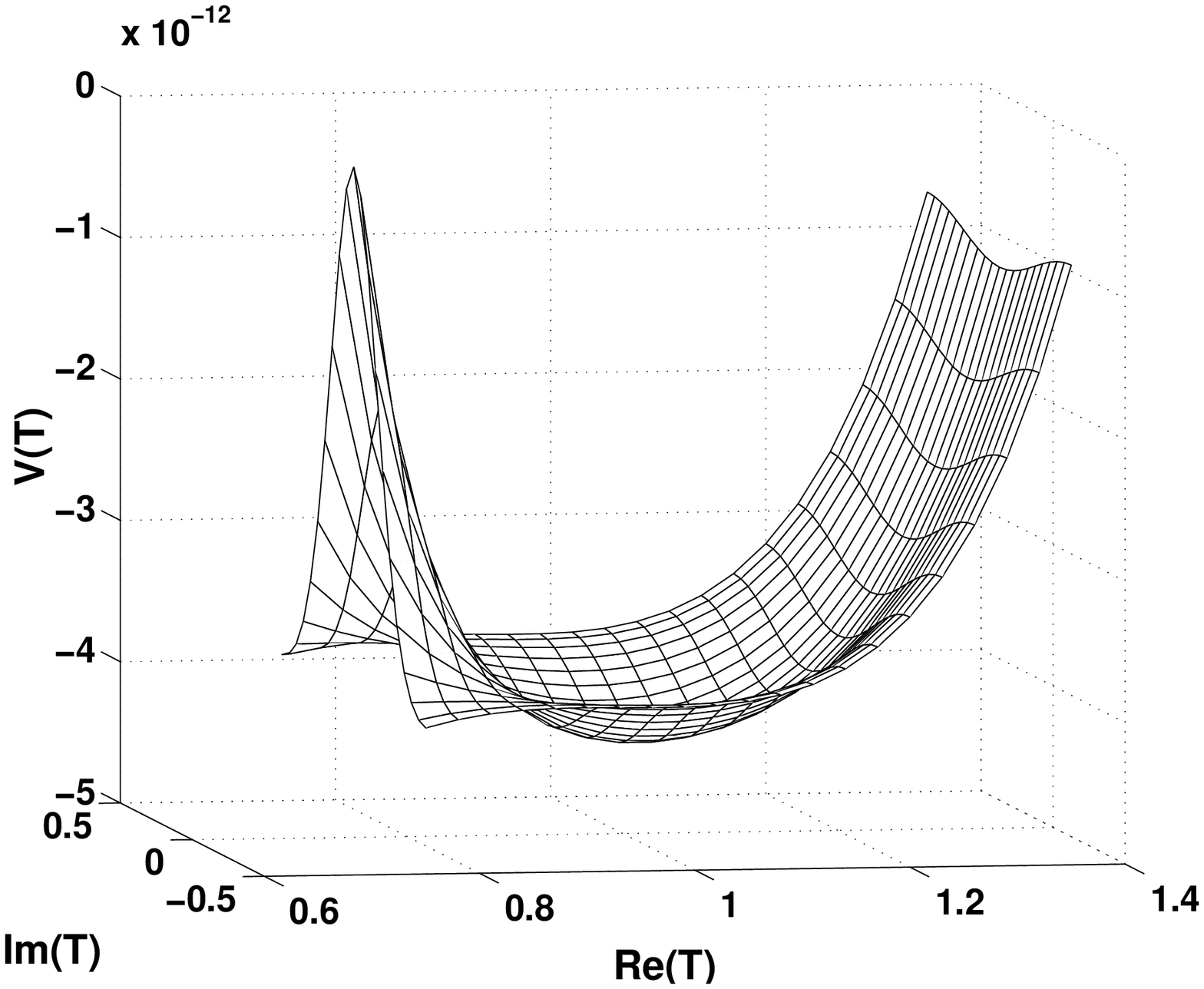}
	\caption{Scalar potential for $m=n=0$, $p(J)=1$, $\beta=-0.3$, 
	$V_{\rm dil}=0$, $\delta_{\rm GS}=15$.}
	\label{fig:VT_FS0d15}
	\end{figure}

\begin{figure}
	\centering
	\includegraphics[width=8.5cm,height=6.5cm]{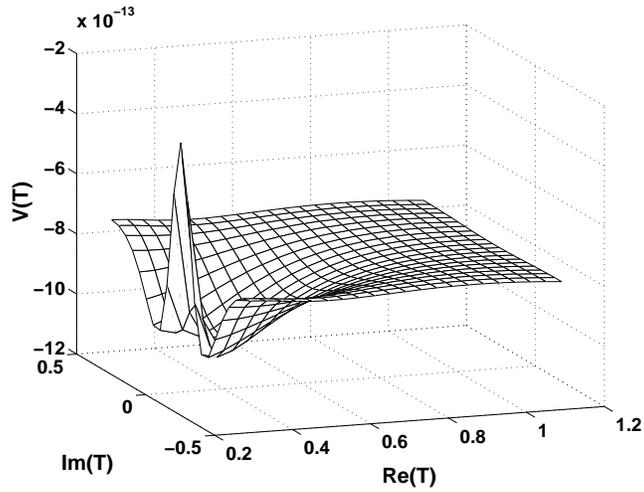}
	\caption{Scalar potential for $m=n=0$, $p(J)=1$, $\beta=-0.3$, 
	$V_{\rm dil}=0$, $\delta_{\rm GS}=-10$.}
	\label{fig:VT_FS0dn10}
	\end{figure}

\begin{figure}
	\centering
	\includegraphics[width=8.5cm,height=6.5cm]{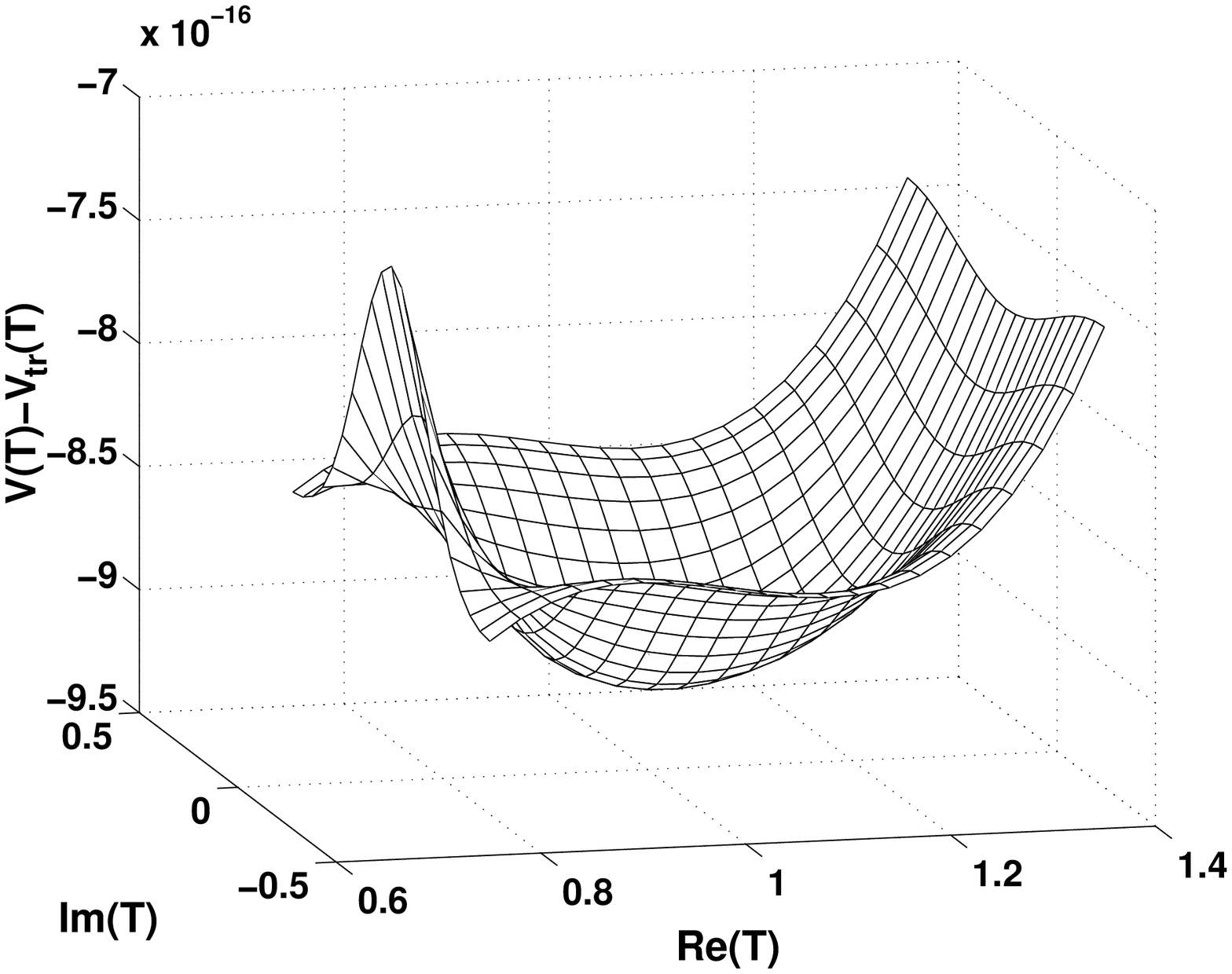}
	\caption{Corrections to the truncated potential for $m=n=0$, $p(J)=1$, 
	$\beta=-0.3$, $V_{\rm dil}=0$, $\delta_{\rm GS}=0$.}
	\label{fig:dV_FS0d0}
	\end{figure}

For $V_{\rm dil}=0$, corresponding to no supersymmetry-breaking in the dilaton sector, the
minima of the scalar potential lie along the real axis. At $\delta_{\rm GS}=0$ the minima are
at approximately $T=0.8$, $T=1.22$ (fig.~\ref{fig:VT_FS0d0}), 
as $\delta_{\rm GS}$ increases the minima approach $T=1$ and merge (fig.\ \ref{fig:VT_FS0d15})
while at negative
$\delta_{\rm GS}$ the minima become more widely separated: at $\delta_{\rm GS}=-10$ minima occur
at $T\simeq 0.4$, $2.5$ (fig.~\ref{fig:VT_FS0dn10}). 
Note that the minima are related by the
modular symmetry $T\mapsto 1/T$; there are also minima at the modular images under
$T\mapsto T+i$, which are not shown. We have also checked numerically that the scalar
potential is an invariant function under the full SL$(2,{\mathbb Z})$ modular group.
The corrections to the truncated approximation (at $\delta_{\rm GS}=0$) are shown in fig.\
\ref{fig:dV_FS0d0}.
As discussed in \cite{me99} they have a $T$-dependence different from that
of the truncated scalar potential, so in principle the corrections could alter the position
of the minima. In this case the effect is not significant for phenomenology since the 
minima will remain on the real axis and at $T$-values of order unity.

\begin{figure}
	\centering
	\includegraphics[width=8.5cm,height=6.5cm]{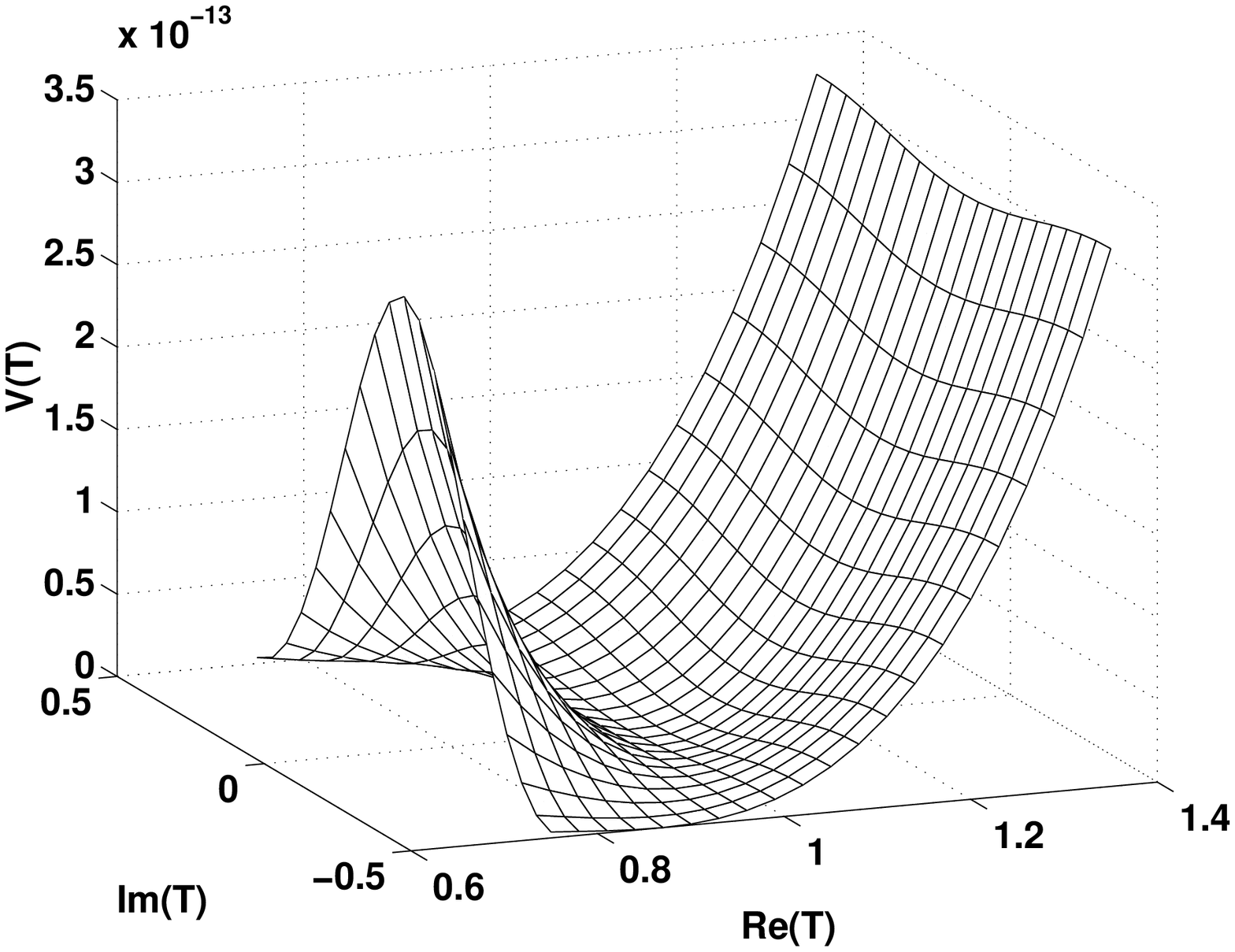}
	\caption{Scalar potential for $m=n=0$, $p(J)=1$, $\beta=-0.3$, 
	$V_{\rm dil}=3$, $\delta_{\rm GS}=0$.}
	\label{fig:VT_FS3d0}
	\end{figure}

When $V_{\rm dil}=3$, the cosmological constant is fine-tuned to zero at the minimum of the 
scalar potential (fig.~\ref{fig:VT_FS3d0}).
In this case, there are two degenerate minima inside ${\mathcal F}$ at $T=1$, $\rho$ 
(where $\rho=e^{i\pi/6}$) with $C'_2=0$ at these points, and the positions of the minima are 
unaffected by changing $\delta_{\rm GS}$. Since the corrections to the truncated approximation 
vanish at $C'_2=0$, in this special case the minima are also unaffected by the corrections.

We also take $V_{\rm dil}=6$, which corresponds to a large, positive cosmological constant. 
The minima are now at $T=\rho$ and its images under modular transformation. As in the case 
where $V_{\rm dil}=3$, the position of the minimum is not changed by taking $\delta_{\rm GS}\neq 
0$, and the effect of the corrections to the truncated approximation is small. We can 
conclude that when the Dedekind eta function is the only modular form arising in the 
threshold corrections, either $T$ is real at the minimum, in which case CP-violating phases
vanish, or $T=\rho$, in which case $F^T$ vanishes and the moduli do not contribute to
supersymmetry-breaking. While this scenario is satisfactory as a solution to the 
supersymmetric CP problem (and, when $F^T=0$, the supersymmetric flavour problem also)
it does not make any characteristic predictions for CP-violating quantities different from
those of the standard model. If there are no other sources of CP violation which can 
generate a CKM phase then the scenario is, of course, ruled out. Neither does it throw light 
on the problems of CP violation in the SM, namely the origin of the cosmological baryon 
asymmetry and the high value of $\epsilon'_K/\epsilon_K$.

\subsection{Results for threshold corrections including $J(T)$}
\subsubsection{$m=1$, $n=0$, $p(J)=1$} \label{sec:FSstatement}
\begin{figure}
	\centering
	\includegraphics[width=8.5cm,height=7cm]{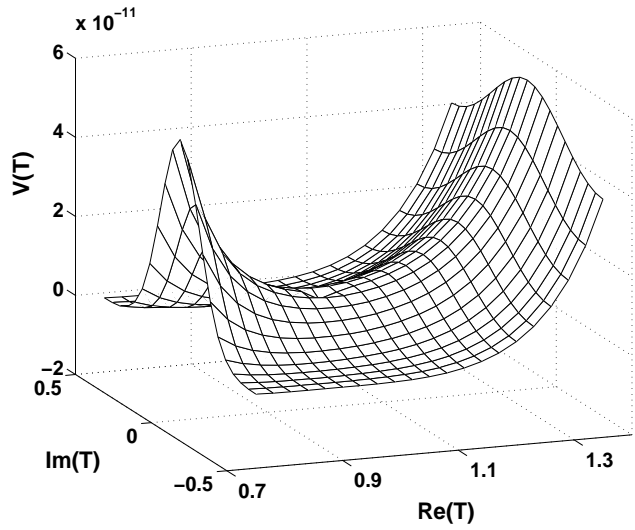}
	\caption{Scalar potential for $m=1,n=0$, $p(J)=1$, $\beta=-0.3$, $V_{\rm dil}=0=
	\delta_{\rm GS}$.}
	\label{fig:VTC_FS0d0}
	\end{figure}

\begin{figure}
	\centering
	\includegraphics[width=8.5cm,height=7cm]{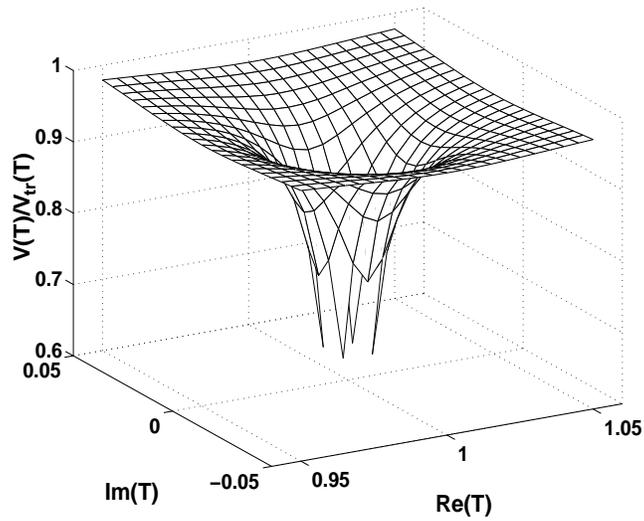}
	\caption{Scalar potential divided by the truncated potential near $T=1$, for the 
	same values of parameters as fig.~\ref{fig:VTC_FS0d0}.}
	\label{fig:VTConVT0}
	\end{figure}

First we take the case when the function $H(T)$ is just proportional to $(J(T)-1)^{1/2}$.
This is in some sense natural, since $J-1\propto(T-1)^2$ near $T=1$, so the square root
remains well-defined near the zero of $H$. However, as discussed in \cite{CveticFILQ}, the 
effective action for the condensate may become ill-defined at some finite radius around
$T=1$, since some string states are supposed to become light here; we also expect the 
truncated approximation to fail badly near $T=1$ since the quantity $x=AC'_2|z_{\rm tr}|^2$ 
which measures the size of deviations from the truncated approximation becomes large
\cite{me99}. We first present the form of the scalar potential for 
$V_{\rm dil}=0=\delta_{\rm GS}$ (fig.~\ref{fig:VTC_FS0d0}).	
Note the ``dimple'' near $T=1$ which is related to the failure of the truncated 
approximation. We can look at this area of the complex $T$ plane in the limit where
$T\rightarrow 1$ and find that while the truncated condensate value $|z_{\rm tr}|$ vanishes as 
the cube root of $(T-1)$, $C_2'$ diverges as $(T-1)^{-2}$, so the truncated approximation 
fails at a certain radius from $T=1$. The effect on the scalar potential as a function of
$T$ is shown in fig.~\ref{fig:VTConVT0}. 
At the values of $T$ where a value for the scalar potential is not plotted, the gaugino 
condensate is destabilized by the $C'_2|z|^6$ term in the scalar potential. Here we appear
to be on a branch of solutions with zero condensate and unbroken supersymmetry 
\cite{me99,Goldberg95} which is not connected to the rest of the surface $V(T)$. If the 
effective action for the gaugino condensate is valid down to $z=0$ then there exists a 
supersymmetric, zero-energy vacuum for all $T$ and it becomes a dynamical and cosmological
question as to how the condensate becomes non-zero \cite{Goldberg95}.

The minimum of the scalar potential is near $T=\rho$: this point is actually a (rather flat) 
{\em maximum} and the minima are close by inside the boundary of ${\mathcal F}$ at 
$T=0.8842 + 0.4844i$  and its images under modular transformations (and the complex 
conjugate of these values). The cosmological constant is negative. In this case 
$\hat{K}^TF^T/m_{3/2}=-0.0389+ 0.0136i$; we use this quantity as a rough measure of the 
amount of supersymmetry-breaking and CP violation that we expect to originate from a 
particular v.e.v.\ of $T$\footnote{For $\delta_{\rm GS}=0$ we have $\hat{K}^TF^T/m_{3/2}=
-3-(T+T^*)(6d\ln \eta(T)/dT-d\ln H(T)/dT)$ (see eqs.~\ref{eq:Atr}--\ref{eq:FSTtr}); 
for non-zero $\delta_{\rm GS}$ the analogous quantity is \newline $\hat{K}^TF^T/m_{3/2}=
-(1-3\delta_{\rm GS}/b_0)(3+6(T+T^*)d\ln \eta(T)/dT)+(T+T^*)d\ln H(T)/dT$ (compare 
\cite{deCarlos:1993pd,Bailin:1997fh} and Eq.~\ref{eq:C2'}).}. This result may be 
compared to the case where $m=0$, $\delta_{\rm GS}=-5$ and the potential is minimised at 
$T=1.503$, which results in $\hat{K}^TF^T/m_{3/2}=-1.172$, a factor of 50 larger. 
Where $T$ is stabilised very near the self-dual point $\rho$ (where $F^T$ vanishes) 
with $m=1$, $F^T$ appears to be much smaller than at a general point inside ${\mathcal 
F}$, so its contribution to CP and flavour violation should be small. But we must ask, 
small relative to what? In order to have phenomenologically reasonable soft breaking 
terms we also require $F^S$ to be non-zero, so that for $F^T$ small we are near the 
dilaton-dominated limit of supersymmetry-breaking. Since $V_{\rm dil}\propto |F^S|^2$ we 
should consider the effect of changing $V_{\rm dil}$ on the stabilization of $T$, as well 
as varying $\delta_{\rm GS}$ and looking at the effects of deviations from the truncated 
approximation. 

We first take $V_{\rm dil}=1.5$ while keeping $\delta_{\rm GS}=0$ and 
$\beta\equiv b_0/(16\pi^2)= -0.3$ fixed: then the minimum is on the unit circle near 
$T=\rho$ at $T=0.8777 + 0.4792i$ and $F^T=0$. For $V_{\rm dil}=3$, the minimum is at 
$T=\rho$, $F^T=0$ and the cosmological constant is tuned to zero.

Next consider changing $\delta_{\rm GS}$: we first take $\delta_{\rm GS}=-15$, $V_{\rm dil}=0$, 
$\beta=-0.3$, in which case the minimum is on the boundary of ${\mathcal F}$ at $T=
0.9058 +0.5i$ and $F^T =-0.1075$ is real at the minimum. Considerations of modular 
invariance \cite{Dent01} indicate that this v.e.v.\ for $T$ will not generate
CP violation, however small deviations from universality may result. For $\delta_{\rm GS}=-15$,
$V_{\rm dil}=3$, $\beta=-0.3$ the minimum is extremely close to $T=\rho$ and $F^T$ vanishes. 
For $\delta_{\rm GS}=15$, $V_{\rm dil}=0$, $\beta=-0.3$ the minimum is on the unit circle at 
$T=0.8873+ 0.4611i$ and $F^T$ also vanishes; we find that this result is virtually unaffected 
by changing $V_{\rm dil}$.

Finally for this case we consider the effect of corrections to the truncated approximation 
on the minimisation of the scalar potential. Since these scale with $|z_{\rm tr}|^2 \propto 
e^{f_{\rm g}/\beta}$, the corrections can be turned on or off by changing $\beta$ (although only 
a finite range of values will result in a phenomenologically reasonable scale of 
supersymmetry-breaking). The truncated approximation corresponds to minimising the scalar 
potential in the limit $|z_{\rm tr}|\rightarrow 0$, {\em i.e.}\/\ $\beta \rightarrow 0^-$, 
while the corrections are maximised for large, negative $\beta$. 

We first take $\beta=-0.1$ to mimic the truncated approximation: for $V_{\rm dil}=0
=\delta_{\rm GS}$ the minimum is at $T=0.8842+0.4844i$ and 
$\hat{K}^TF^T/m_{3/2}=-0.0388 + 0.0134i$: differing only very slightly from the result 
at $\beta=-0.3$. For $\delta_{\rm GS}=0$, $V_{\rm dil}=3$ the minimum is again at
$T=\rho$. Next we take the extreme large value $\beta=-0.9$, keeping $V_{\rm dil}=0
=\delta_{\rm GS}$, which results in a minimum at $T= 0.8844+0.4847i$ with $\hat{K}^TF^T/m_{3/2}
= -0.0403 + 0.0136i$. We see in this case that corrections to the truncated 
approximation do not much change the position of the minimum.

\subsubsection{$m=n=0$, $p(J)=J-1$} \label{sec:pisJ-1}
This ansatz, equivalent to $m=2$, $n=0$, $p(J)=1$ reproduces the desirable feature of the 
previous case of a small or zero contribution of the $T$ modulus to CP violation and 
nonuniversality, without the ``hole'' where the gaugino condensate is destabilised. 
However, due to the zero of $J$ at $T=1$, the condensate vanishes smoothly at this point,
an equally undesirable scenario for phenomenology! As for the case $m=1$, $n=0$, the 
phenomenologically interesting minima are near $T=\rho$.
We start with the same set of parameters, $V_{\rm dil}=0=\delta_{\rm GS}$ and $\beta=-0.3$, for
which the minimum is inside the fundamental domain at $T=0.8754 + 0.4921i$ and $\hat{K}^T
F^T/m_{3/2} =-0.0199 + 0.0071i$ and the cosmological constant is negative. For $V_{\rm dil}=3$ 
there is a minimum on the unit circle at $T=0.8721 + 0.4894i$ with $F^T=0$, and $T=\rho$ 
also appears to be a minimum. 

In this case corrections to the truncated approximation 
may be significant. As we increase $\beta$, and thus the size of the corrections, the 
minimum inside the fundamental domain for $V_{\rm dil}=0$ moves towards the line Im$(T)=1/2$,
although only slowly. We may compare the values of $\hat{K}^TF^T/m_{3/2}$ at the minmum: 
for $\beta=-0.3$ it is 0.87539 + 0.49207i and for $\beta=-0.9$ we have $\hat{K}^TF^T/
m_{3/2}= 0.87547 + 0.49215i$. The effect is small, which follows from the fact that the
corrections tend to be smallest around the minimum of the potential.

For $\delta_{\rm GS}=-15$, $V_{\rm dil}=0$, $\beta=-0.3$ the modulus is stabilised at 
$T= 0.8857 +0.5i$ and $\hat{K}^TF^T/m_{3/2} = -0.0538$. At the larger value $\beta=-0.7$ 
the minimum remains on the line Im$(T)=1/2$, so again the corrections do not seem to have 
a large effect. For $\delta_{\rm GS}=+15$, $V_{\rm dil}=0$, $\beta=-0.3$ the minimum is on the 
unit circle at $T=0.8774+ 0.4796i$ such that $F^T=0$: this conclusion is also unchanged by
increasing $\beta$ and thus the corrections to truncation.

\subsubsection{$m=0$, $n=1,2,\ldots$, $p(J)=1$}
In this case the scalar potential is minimised at $T=1$ with $F^T=0$, regardless of the 
values of other parameters. Near $T=\rho$ the same phenomenon occurs as in the case of 
$m=1$, $n=0$, $p(J)=1$ near $T=1$: at some finite radius away from the fixed point the
truncated approximation fails and a non-zero value for the condensate becomes unstable.
If we set $m=0$, $n=3$, $P(j)=1$, which is equivalent to $m=n=0$, $p(J)=J$, the 
supersymmetry-breaking minimum is again at $T=1$ and the condensate goes smoothly to zero 
as $T\rightarrow \rho$.

\subsubsection{$m=1,2$, $n=1,2$, $p(J)=1$} \label{sec:m12n12}
In this case there is a 
region around both fixed points $T=1$, $T=\rho$ where the condensate is destabilised or
goes continuously to zero. 
	\begin{figure}
	\centering
	\includegraphics[width=8.5cm,height=7cm]{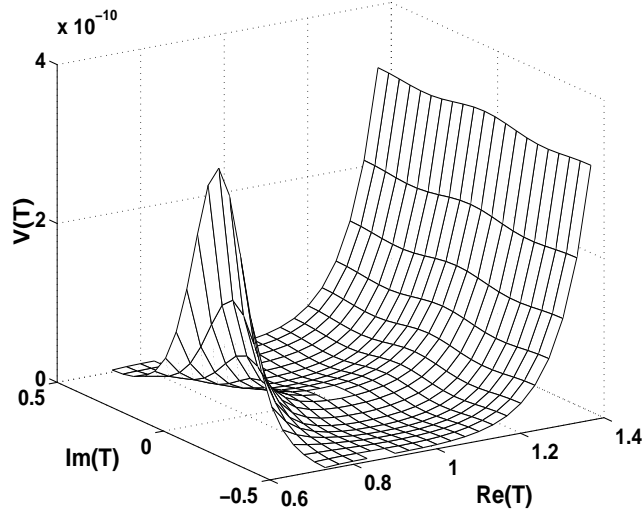}
	\caption{Scalar potential for $m=n=1$, $p(J)=1$, $\beta = -0.3$, $V_{\rm dil}=0=
	\delta_{\rm GS}$.}
	\label{fig:VTG_d0}
	\end{figure} 
For $m=n=1$ and the parameter values $V_{\rm dil}=0$, $\delta_{\rm GS}=0$, $\beta=-0.3$, 
there are minima on the unit circle at $T=0.9713+ 0.2378i$ (and at the complex conjugate 
value) with $F^T$ vanishing (fig.~\ref{fig:VTG_d0}). Again, the minimisation is robust to 
changes in parameters, in that the minimum remains on the unit circle with $F^T=0$ when 
$V_{\rm dil}$, $\delta_{\rm GS}$ and $\beta$ are changed within reasonable limits.

\subsection{Interpretation of the results}\label{sec:interp}
The main results from minimization in any particular case are the value of $T$ at the 
minimum and the size of the quantity $\hat{K}^TF^T/m_{3/2}$, which measures the contribution
of the modulus to soft supersymmetry-breaking terms. Since the phase of $F^T$ is not 
modular invariant, and the phases of the soft terms receive additional contributions from
the $T$-dependence of the Yukawa couplings (which we have not explicitly calculated), we 
cannot give unambiguous measures of CP violation in the soft terms. However the value
of $\langle T\rangle$ should allow us to diagnose whether CP is broken in the low-energy
theory, and the size of $F^T$ allows us to estimate how far we are from the dilaton- or
moduli-dominated limits. The results presented seem to indicate that the most common patterns
of supersymmetry-breaking are close to the dilaton-dominated limit, except when 
$V_{\rm dil}=0$ and $\eta(T)$ is the only modular form appearing in the threshold 
corrections: for various values of parameters, we obtain $F^T$ vanishing or small compared 
to the scale of the gravitino mass, and when $F^T$ is nonzero the v.e.v.\ of $T$ is such 
that its contribution to supersymmetry-breaking may be either CP-conserving or (if 
$\langle T\rangle$ lies in the interior of $\mathcal{F}$) CP-violating.

\subsection{Singular points in the effective theory}
We noted several times, in the case where the threshold corrections involved $J(T)$, points
in the $T$ plane where the effective potential apparently became discontinuous, when the 
gaugino condensate was allowed to be dynamical (see in particular \ref{sec:FSstatement}). This
was due to divergences in the threshold corrections, which can be interpreted as the effects
of new light states appearing \cite{CveticFILQ}. Then, rather than a transition to another 
branch of the theory with zero condensate, one should conclude that the effective theory is 
breaking down. In the case of new massless matter fields, the correct effective
theory at these points may look something like supersymmetric QCD, which has a runaway 
vacuum and equally unappealing phenomenology. 

In the usual approximation of a nondynamical condensate, such points do not appear 
pathological, and simply give extrema or zeros of the scalar potential at which the effective 
superpotential vanishes. The apparent smooth behaviour is due to a conspiracy between a 
vanishing gaugino condensate and a divergent factor proportional to 
$W^{-1}\partial W/\partial T$ \cite{me99} which in the dynamical condensate case would signal 
the breakdown of the effective theory. Many of the minima quoted in \cite{KhalilLM} are of 
this type; while this fact does not alter the negative phenomenological conclusions, it 
should be noted that since the effective theory is unlikely to be valid at such points, the 
meaning of results derived from it is unclear.

\section{Further reflections}\label{sec:further}
The existence of a viable phenomenology of CP violation in Yukawa couplings and soft terms 
depends largely on the dynamics of the $T$ modulus in these models. The main contributions 
that the dilaton dynamics should make are to ensure the correct unified gauge coupling and,
in the case where $F^S\neq 0$, to contribute to soft terms in such a way as to alleviate the 
problems of nonuniversality and small gaugino masses in the moduli-dominated limit. We 
assumed, with \cite{Bailin:1998iz,Bailin:1998xx}, that the dilaton plays essentially no 
role in the minimization in the $T$-direction. The results of \cite{KhalilLM} show this
to be a valid assumption for the racetrack and S-dual mechanisms, for which the minimum in
$S$ is found algebraically by the vanishing of $G^S$, where $G\equiv K+\ln |W|^2$. Thus, 
while knowledge of the dilaton dynamics is needed to construct a full model, it is less 
important if one is mainly concerned with CP violation by $\vev{T}$. The nontrivial minima 
obtained here for the cases ($m=1$, $n=0$), ($m=2$, $n=0$) and $m=n=1$ are reproduced in 
\cite{KhalilLM}, up to a modular transformation (see their Tables 2 and 3).
It is possible that a nonzero $F^S$ with a complex phase could produce important 
effects in the soft terms; for example, using racetrack stabilization with 
$\delta_{\rm GS}\neq 0$ \cite{KhalilLM}, although the complex values of $\vev{F^S}$ here 
appear to be due to mixing with complex values of $\vev{F^T}$ and $\vev{T}$. 
Since one cannot be confident that the correct dilaton stabilization scheme has been 
discovered, a complex $F^S$ should be allowed, but we are not really in a position to predict 
its value. 

We already mentioned in \cite{Dent01} the possibility that $\vev{S}$ might be the 
source of CP violation, in cases where the value of $\vev{T}$ is CP-conserving, {\em i.e.}\/\
on the boundary of ${\mathcal F}$. We concluded that this was ruled out since the dilaton 
couplings are universal and cannot give the required flavour structure to reproduce observed 
signals of CP violation while respecting EDM bounds. This anticipated the ``important remark'' 
made in \cite{KhalilLM} that fixed-point values of $\vev{T}$ cannot explain the observations,
regardless of the dilaton dynamics.

However, there are differences between our results and \cite{KhalilLM},
namely in the case with nonperturbative dilaton K\"ahler potential. The property claimed in 
\cite{Casas96} of giving $F^S\neq 0$ only survives in a single case, $m=n=0$, when
the minimisation in the $S$-direction is carried out explicitly \cite{KhalilLM}. This occurs 
because all the minima have $T$ at the fixed points, where $W$ vanishes unless $m=n=0$, and 
the modular invariant functions chosen vanish at the minima. Then $F^S$ must vanish, being 
proportional to $|W|$. The discussion of the relation between $V_{\rm dil}$ and $F^S$ in our 
thesis (section \ref{sec:FSstatement}) was slightly incomplete since we did not consider 
the case that $W$ vanished at the minimum. 
Technically, $|W|^2 V_{\rm dil}\propto |WG^S|^2\propto |F^S|^2$, where $G\equiv K+\ln |W|^2$. 
Thus, while in the racetrack and $S$-dual cases both $G^S$ and $F^S$ vanish, corresponding to 
setting $V_{\rm dil}=0$ (which was our default value), for K{\" a}hler stabilization
the $F$-term vanishes if $W=0$, even though $G^S$ is nonzero.

In terms of our treatment this means that the K{\" a}hler stabilization case should 
correspond to a nonzero $V_{\rm dil}$, which we did in many cases include in the analysis. 
The effect is indeed to change the shape of the potential $V(T)$, although even with 
$V_{\rm dil}\neq 0$ we found different minima from those of \cite{KhalilLM}, 
including some not at fixed points, at which $|W|$ likely does not vanish. 
Specifically, for $m=1$, $n=0$ we found a minimum near the fixed point $T=\rho$ with 
$F^T=0$ for $\delta_{\rm GS}=0$ and nonzero $G^S\propto V_{\rm dil}^{1/2}$, and a minimum 
on the line Im$\,T=1/2$ with $F^T\neq 0$ for $\delta_{\rm GS}=15$ and a wide range of 
values of $G^S$ (Section~\ref{sec:FSstatement}). For $m=2$, $n=0$, $\delta_{\rm GS}=0$ and 
nonzero $G^S$ we found minima on the unit circle close to $T=\rho$ (Section~\ref{sec:pisJ-1}) 
\footnote{The vanishing of $F^T$ for $T$ on the unit circle is due to a general property of the 
modular functions considered, and does not imply that $\vev{W}=0$ at all such points.}. For 
$m=n=1$ we found minima with $T$ on the unit circle which remained on the unit circle (avoiding 
fixed points) for various values of $G^S$ (Section~\ref{sec:m12n12}); a similar result
was obtained in \cite{Bailin:1998iz} for slightly different values of parameters.

If the dilaton v.e.v.\ at the minimum of the potential is not sensitive to the value of $T$, 
then the assumption of constant values of $S$ and of $G_S$ is justified; there is little to 
choose between parameterising the dynamics by the constants $\vev{y}$, $V_{\rm dil}$ (or $G^S$) 
and $P'(y)$, or by the $d$, $p$, $b$ of \cite{Casas96}, so far as stabilizing 
$T$ is concerned. There is currently no way to calculate either set of parameters. Our approach 
is more flexible, since we can consider the effect of varying $V_{\rm dil}$ (equivalently
$G_S$) from zero to any finite value, while the alternative is to be tied to zero, for the
racetrack or S-dual mechanisms, or the values produced by a particular form of the dilaton 
K{\" a}hler potential. The only case where the parameterization by constant $\vev{y}$, 
$V_{\rm dil}$ and $P'(y)$ may be invalid is if the dilaton stabilization has a nontrivial 
dependence on the value of $T$, thus altering the shape of $V(T)$ still further by a 
non-constant prefactor $|z|^6\propto e^{24\pi^2 y/b_0}$. Such a nontrivial interplay 
between the minimisations in the $S$- and $T$-directions may account for the difference of 
results between the two approaches. 

Further investigation along the lines of \cite{KhalilLM} could easily resolve this question, 
by finding a nonzero variation in $S_{|{\rm min}}$ (or $y_{|{\rm min}}$, in the case
of nonzero $\delta_{GS}$), as $T$ is varied by hand. 
If minima with nonvanishing superpotential and $\vev{T}$ inside the fundamental domain can be 
found with the K{\" a}hler stabilization mechanism it would be phenomenologically important, 
since $F^S \neq 0$ is crucial for viable superpartner spectra, and conversely 
values of $\vev{T}$ on the boundary cannot be the source of CP violation. 

\section*{Acknowledgements}
The author acknowledges the invaluable supervision of David Bailin at the time when much of 
this work was carried out, and a helpful correspondence with Stephen Morris. This research was 
also supported in part by DOE Grant DE-FG02-95ER40899 Task G.
\subsection*{Authorial note}
Much of the material in this paper, specifically Sections 
\ref{sec:soft_withU}--\ref{sec:interp}, appeared in the author's D.~Phil.\ thesis submitted 
at the University of Sussex in September 2000.



\end{document}